\documentclass{jpp}
\usepackage{graphicx}
\usepackage{siunitx}
\usepackage{ulem}

\usepackage[utf8]{inputenc}
\usepackage[T1]{fontenc}
\usepackage{amsmath}
\usepackage{tabularx}
\usepackage{booktabs}
\newcolumntype{C}{>{\centering\arraybackslash}X}
\usepackage[colorlinks=true, citecolor=blue, linkcolor=blue, urlcolor=blue]{hyperref}

\shorttitle{Critical velocity-space mode scalings in Landau damping}
\shortauthor{Guberman, Coughlin, and Joglekar}

\title{Critical velocity-space mode scalings in linear and nonlinear Landau damping for the Vlasov--Poisson system}

\author{Noah K. Guberman\aff{1,2}\corresp{\email{noahgub@umich.edu}}, J. Coughlin\aff{3}, \and A. S. Joglekar\aff{1,2,3}}
\affiliation{\aff{1}Ergodic LLC, Seattle, WA \\
\aff{2}Laboratory for Laser Energetics, University of Rochester, Rochester, NY \\
\aff{3}Pasteur Labs, Brooklyn, NY}

\begin{document}

\maketitle

\begin{abstract}
The velocity-space resolution required to accurately simulate kinetic phenomena in the 1D-1V Vlasov--Poisson system is generally not known a priori. In this work, we determine the upper bound on the resolution requirement for linear and nonlinear Landau damping mediated by collisional diffusion, deriving analytical scalings for the critical Fourier and Hermite velocity-space mode numbers using a unified cascade-balance argument. The resulting scalings depend on the bounce frequency $\omega_b$, wavenumber $k\lambda_D$, and electron-electron collisional frequency $\nu$. We validate these predictions against an ensemble of 800 Vlasov--Fokker--Planck simulations, finding strong agreement with the predicted $\omega_b$ and $\nu$ dependencies.
\end{abstract}

\section{Introduction}
\label{sec:intro}

Landau damping is the collisionless transfer of energy from an electrostatic wave to resonant particles, a process that preserves the information of the system while damping its macroscopic fields. In the kinetic description, this manifests as phase mixing, whereby the distribution function $f(t,x,v)$ develops increasingly fine structure in phase space \citep{van_kampen_theory_1955, case_plasma_1959}. In a strictly collisionless system, phase mixing results in information cascading indefinitely towards higher velocity-space modes, rendering the damping theoretically reversible \citep{schekochihin_astrophysical_2009}. However, physical irreversibility requires a mechanism to arrest this cascade and smooth the distribution function. In the presence of an arbitrarily small, non-zero collisional frequency, collisional diffusion arrests the cascade. With linear phase mixing (filamentation), the ballistic streaming of unperturbed particles at differing velocities shears the distribution function into increasingly fine scales \citep{bernstein_exact_1957}; with nonlinear phase mixing, the trapping of resonant particles within the wave's potential well replaces linear phase mixing as the driver of the cascade through the formation of a fine separatrix layer at the boundary of the phase-space vortex \citep{oneil_collisionless_1965}.

Quantifying the velocity-space mode number at which collisional diffusion arrests phase mixing is of practical importance for numerical simulation. Resolving all scales down to the critical mode sufficiently captures the long-time behaviour of Landau-damped waves in continuum Vlasov codes, while over-resolving wastes computational resources. The critical mode provides an upper bound on this requirement, whereas the minimum resolution required to accurately reproduce a specific kinetic observable remains an open question. %, and the two representations exhibit different cascade dynamics and therefore different critical mode scalings.

Both Fourier and Hermite spectral representations of velocity-space are widely used in computational and theoretical applications. The utility of decomposing the distribution function onto Hermite polynomials is well established; truncating the Vlasov equation to an expansion in the first $n$ Hermite polynomials is equivalent to evolving the first $n$ velocity moments of the distribution function, thereby connecting the kinetic description and fluid model of a plasma \citep{armstrong_numerical_1967,grant_fourierhermite_1967}. This approach is of particular relevance to gyrokinetics and electrostatic turbulence research \citep{hammett_developments_1993, watanabe_kinetic_2004, zocco_reduced_2011, hatch_transition_2013} and is likewise useful in the 1D-1V Vlasov--Poisson system for the same reasons \citep{parker_fourierhermite_2015}. A key open question in this context is how many Hermite modes are needed to faithfully capture the kinetic physics in the 1D-1V system. The critical mode scalings presented in this paper provide a direct, parameter-dependent upper bound on the necessary Hermite resolution: for given bounce frequency $\omega_b$, wave number $k\lambda_D$, and electron-electron collisional frequency $\nu$, modes beyond the critical Hermite mode carry negligible information and can be truncated.

The critical mode scaling in the linear, weakly collisional regime is well established. The $s_c \propto (k/\nu v_{\rm th}^2)^{1/3}$ scaling of the Fourier velocity wavenumber at which collisional diffusion arrests linear phase mixing has been derived by numerous authors \citep{su_collisional_1968, johnston_dominant_1971, auerbach_collisional_1977, callen_coulomb_2014, catto_weakly_2025}. The corresponding linear Hermite cutoff $m_c \propto (kv_{\rm th}/\nu)^{2/3}$ follows from the same balance expressed in the Hermite basis \citep{zocco_reduced_2011,kanekar_fluctuation-dissipation_2015}. In the nonlinear regime, however, the picture is considerably less settled. \citet{schekochihin_phase_2016} derived the nonlinear critical Hermite mode scaling in the context of gyrokinetic turbulence in a magnetized plasma. Several studies have also examined weakly collisional nonlinear Landau damping and the role of trapping in saturating the wave \citep{parker_fourierhermite_2015, catto_weakly_2025}. To our knowledge, the critical velocity-space mode number at which collisional diffusion arrests the trapping-related cascade has not yet been characterised. This gap is what the present work addresses.

In this paper, we derive the critical Fourier and Hermite velocity-space mode numbers for both linear and nonlinear Landau damping in a 1D-1V plasma governed by the Vlasov--Fokker--Planck (VFP) equation with a Lenard--Bernstein collision operator. The linear scalings recover established results from a unified cascade-balance argument. We obtain the nonlinear scalings by replacing the cascade timescale with the particle bounce (trapping) time, yielding predictions for the dependence of the critical mode on $\omega_b$, $k\lambda_D$, and $\nu$. We verify all four scalings against an ensemble of \texttt{ADEPT} simulations spanning a range of $a_0$, $k\lambda_D$, and $\nu$, where $a_0$ is the driver amplitude.

This paper is organised as follows. In Section~\ref{sec:critical-mode-scaling} we derive the critical Fourier and Hermite mode scalings analytically for both the linear and nonlinear regimes. While the derivations in the linear case are well known, we include them here for completeness. Section~\ref{sec:model} describes the simulation setup and the numerical algorithm used to extract critical modes from the computed distribution functions. Section~\ref{sec:results} presents the power-law fits and their comparison with theory. Conclusions are given in Section~\ref{sec:conclusion}.

\section{Critical mode scaling}
\label{sec:critical-mode-scaling}

\begin{figure*}
    \centering
    \includegraphics[width=\textwidth]{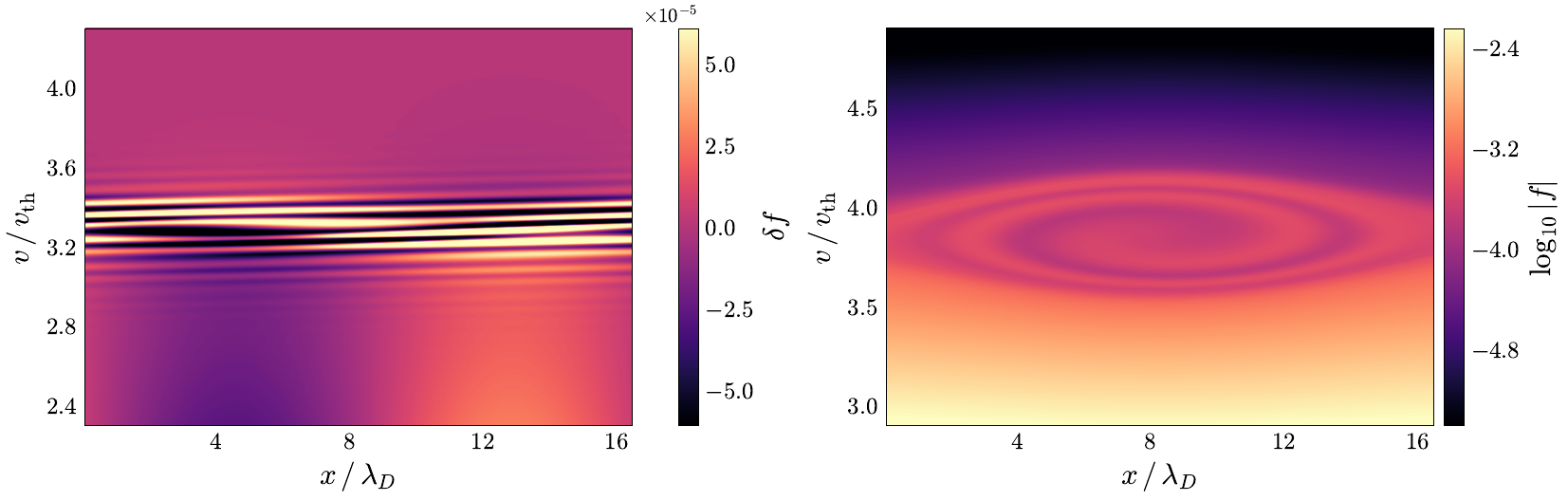}
    \caption{\textit{Left}: linear (small-amplitude) phase mixing showing filamented structure in the mean-subtracted distribution function at $t = 800$ $\omega_p^{-1}$. \textit{Right}: nonlinear (large-amplitude) phase mixing showing a pronounced vortex and its separatrix boundary in the total distribution function at $t = 1200$ $\omega_p^{-1}$.}
    \label{fig:phase-space}
\end{figure*}

Figure~\ref{fig:phase-space} shows representative snapshots of the resonant region of the distribution function in the two regimes studied here. The linear case exhibits smooth, filamented structure characteristic of linear phase mixing, whereas the nonlinear case displays a phase-space vortex with a distinct separatrix boundary layer characteristic of particle trapping. These two qualitatively distinct velocity-space dynamics motivate separate derivations of the critical mode in each regime.

Throughout this paper, waves are excited by a finite-time external ponderomotive driver of the form $E_D(t,x) = a_0\sin(kx - \omega t)$, where $a_0$ is the driver amplitude normalised to $m_e\omega_pv_{\rm th}/e$. After the driver is switched off, the plasma supports a free electron plasma wave (EPW) whose potential amplitude $\phi_0$ is determined by the integrated plasma response to the drive; in the parameter regimes considered here, $e\phi_0/m_ev_{\rm th}^2$ scales with $a_0$. The driving protocol and the resulting amplitude scaling are described in \citet{joglekar_machine_2023}; for the purposes of the scaling derivations that follow, $a_0$ therefore also characterises the post-drive wave amplitude.

The VFP equation is normalised using characteristic plasma scales. We define the dimensionless variables: $\hat{t} = t\omega_p$, $\hat{v} = v/v_{\rm th}$, $\hat{x} = x/\lambda_D$, $\hat{E} = E/E_0$, and $\hat k = k\lambda_D$, where $v_{\rm th} = \sqrt{T_e/m_e}$, $\lambda_D = v_{\rm th}/\omega_p$, $E_0 = m_e\omega_p v_{\rm th}/e$. The distribution function is normalised as $\hat{f} = f(v_{\rm th}/n_0)$.

In a 1D-1V system with small-angle Coulomb collisions, the distribution function $f(t,x,v)$ is governed by the dimensionless VFP equation,
\begin{equation}
    \partial_{\hat{t}} \hat{f} + \hat{v} \partial_{\hat{x}} \hat{f} + \hat{E} \partial_{\hat{v}} \hat{f} = \hat{C}[\hat{f}],
    \label{eq:VFP}
\end{equation}
where $\hat{C}[\hat f]$ is the dimensionless Lenard-Bernstein collision operator \citep{lenard_plasma_1958, dougherty_model_1964},
\begin{equation}
    \hat{C}[\hat{f}] = \hat{\nu} \partial_{\hat{v}}\left( \hat{v}\hat{f} + \partial_{\hat{v}} \hat{f} \right),
    \label{eq:LB}
\end{equation}
where $\hat \nu = \nu/\omega_p$. For simplicity of notation, we drop the hat accents in the derivations that follow in this section, and all quantities should be assumed to be dimensionless unless otherwise stated.

\subsection{Linear critical Fourier mode}
\label{sec:linear-fourier-mode}

We assume a small perturbation in the distribution function about a background Maxwellian,
\begin{equation}
    f(t,x,v) = f_0(v) + f_1(t,x,v).
    \label{eq:distribution-function}
\end{equation}
By neglecting collisions for the short-time ballistic evolution and retaining only first-order terms, substituting this expansion into Eq.~\ref{eq:VFP} yields the linearized, collisionless Vlasov equation
\begin{equation}
    (\partial_t + v\partial_x)f_1 = - E(x,t) \partial_v f_0.
    \label{eq:VFP-at-f0+f1}
\end{equation}
We integrate along the unperturbed orbit $x(t') = x - v(t-t')$ from $t'=0$ to $t'=t$,
\begin{equation}
    f_1(t,x,v) = f_1(0,x-vt,v) - (\partial_v f_0)\int_0^t \mathrm{d}t'\,E\left(t',x-v(t-t')\right).
    \label{eq:orbit-integral}
\end{equation}
We consider a Fourier mode $e^{ikx}$ such that $f_1(t,x,v) = \hat f_1 (t,k,v)e^{ikx}$ and $E(t,x)=\hat E(t,k)e^{ikx}$. The spatial shift becomes a phase factor $\exp\left(-ikv(t-t')\right)$ and Eq.~\ref{eq:orbit-integral} becomes
\begin{equation}
    \hat f_1(t,k,v) = \hat f_1(0,k,v)e^{-ikvt} - (\partial_vf_0)\int_0^t\mathrm{d}t'\,\hat E(t',k)e^{-ikv(t-t')}.
\end{equation}
Assuming $\hat E(t,k)$ is a monochromatic field and can be expressed as $E_0e^{-i\omega t}$, the integral reduces to
\begin{equation}
    \int_0^t\mathrm{d}t'\, E_0e^{-i\omega t'}e^{-ikv(t-t')} = \frac{iE_0}{\omega-kv}\left(e^{-i\omega t}-e^{-ikvt}\right).
\end{equation}
Hence, the full transformed solution to Eq.~\ref{eq:VFP-at-f0+f1} is
\begin{equation}
    \hat f_1(t,k,v) = \hat f_1(0,k,v)e^{-ikvt} - \frac{iE_0}{\omega-kv}(\partial_v f_0)\left(e^{-i\omega t}-e^{-ikvt}\right).
\end{equation}
Equivalently,
\begin{equation}
    \hat f_1 = -\frac{iE_0}{\omega-kv}(\partial_v f_0)\,e^{-i\omega t} + \left[\hat f_1(0) + \frac{iE_0}{\omega-kv}\partial_v f_0\right]\,e^{-ikvt}.
\end{equation}
The $e^{-i\omega t}$ term on the right is the Landau response of the plasma, oscillating at frequency $\omega$. The $e^{-ikvt}$ term oscillates at $kv$ and is responsible for the development of increasingly fine oscillations in velocity-space \citep{van_kampen_theory_1955, case_plasma_1959}. Importantly, this term represents the dominant long-time behaviour, since the first term undergoes Landau-damping. Therefore, $\hat f_1 \propto e^{-ikvt}$ satisfies the homogeneous version of Eq.~\ref{eq:VFP-at-f0+f1} and carries a dimensionless velocity-space wavenumber
\begin{equation}
    s(t) = kt.
    \label{eq:linear-dependence}
\end{equation}
The Lenard-Bernstein operator (Eq.~\ref{eq:LB}) acts diffusively at small scales. For $f(v) \propto e^{-isv}$,
\begin{equation}
    C[f] \propto \nu (1 - isv)f % friction
    - \nu s^2 f. % diffusion
    \label{eq:LB-at-f}
\end{equation}
The friction term $\nu(1 - isv)f$ is negligible compared to the diffusion term $-\nu s^2 f$ when $s\gg1$. Since collisional diffusion arrests linear phase mixing at large $s$, we can safely neglect the friction term, whereby Eq.~\ref{eq:LB-at-f} yields the damping rate
\begin{equation}
    \gamma_{\rm coll}(s) \propto \nu s^2,
    \label{eq:damping-rate}
\end{equation}
and subsequently the collisional damping time,
\begin{equation}
    t_{\rm coll}(s) \propto \frac{1}{\nu s^2}.
    \label{eq:damping-time}
\end{equation}
Since $s$ grows linearly in time due to streaming (Eq.~\ref{eq:linear-dependence}), the time it takes phase mixing to populate a given $s$ is
\begin{equation}
    t_{\rm cas}(s) \propto \frac{s}{k}.
    \label{eq:cascade-time}
\end{equation}
Equating Eq.~\ref{eq:damping-time} and Eq.~\ref{eq:cascade-time} gives the linear critical Fourier mode
\begin{equation}
    s_c \propto \left( \frac{k}{\nu} \right)^{1/3}, \label{eq:crit-k}
\end{equation}
at which collisional diffusion balances linear phase mixing. In the interest of expressing Eq.~\ref{eq:crit-k} in terms of the familiar product $k\lambda_D$ and to emphasize that $s_c$ is dimensionless, we use the equivalent expression
\begin{equation}
    s_c \propto \left( \frac{k\lambda_D}{\hat \nu}\right)^{1/3},
    \label{eq:critical-wave-number-2}
\end{equation}
for the remainder of this work when referring to the linear critical Fourier mode scaling.

\subsection{Linear critical Hermite mode}

The derivation we present for the linear critical Hermite mode scaling is similar to that of \citet{schekochihin_phase_2016}. We begin by introducing the normalised Hermite polynomials provided by the Rodrigues formula,
\begin{equation}
    \hat H_m(v) = \frac{(-1)^m}{\sqrt{m!}}e^{v^2/2}\frac{d^m}{dv^m}\left( e^{-v^2/2}\right).
    \label{eq:rodrigues}
\end{equation}
These polynomials are orthogonal with respect to the Gaussian weight
\begin{equation}
    \frac{1}{\sqrt{2\pi}} \int_{-\infty}^{\infty}\mathrm{d}v\, \hat H_m(v) \hat H_n(v) e^{-v^2/2} = \delta_{mn}.
\end{equation}
The Hermite transform operator is defined as
\begin{equation}
    \mathcal{H}_m\left[f(t,v)\right] \equiv \tilde f(t,m) \equiv \frac{1}{\sqrt{2\pi}} \int_{-\infty}^\infty \mathrm{d}v\,f(t, v)\hat{H}_m(v) e^{-v^2/2}.
    \label{eq:hermite-transform-operator}
\end{equation}
The ballistic piece of the perturbed distribution function solves $(\partial_t + v\partial_x)\hat f_1 = 0$ (see section~\ref{sec:linear-fourier-mode}). Restricting to a single spatial Fourier mode $e^{ikx}$, this reduces to
\begin{equation}
    \partial_t f_k(t,v) + ikv\, f_k(t,v) = 0.
    \label{eq:pre-sol}
\end{equation}
Applying the Hermite transform operator to Eq.~\ref{eq:pre-sol} yields
\begin{equation}
    \frac{1}{\sqrt{2\pi}} \int_{-\infty}^{\infty}\mathrm{d}v\, \partial_t   f_k(t,v) \hat H_m(v) e^{-v^2/2} + \frac{1}{\sqrt{2\pi}} \int_{-\infty}^{\infty} \mathrm{d}v\, \left( ikv  f_k(t,v) \right) \hat H_m(v) e^{-v^2/2} = 0.
    \label{eq:hermite-vfp}
\end{equation}
The first term evaluates to $\partial_t \tilde f(t,m)$. The second term can be written as
\begin{equation}
    ik \frac{1}{\sqrt{2\pi}} \int_{-\infty}^{\infty} \mathrm{d}v \,  f_k(t,v) \Big[ v \hat H_m(v) \Big] e^{-v^2/2},
    \label{eq:second-term-vlasov-hermite}
\end{equation}
to make use of the recurrence relation
\begin{equation}
    v \hat H_m(v) = \sqrt{m+1}\hat H_{m+1}(v) + \sqrt{m}\hat H_{m-1}(v).
    \label{eq:hermite-recurrence-relation}
\end{equation}
Substituting the recurrence relation into the bracketed term in Eq.~\ref{eq:second-term-vlasov-hermite}, we obtain
\begin{align}
    \frac{ik}{\sqrt{2\pi}}\left[\int_{-\infty}^\infty \mathrm{d}v\, f_k(t,v)\sqrt{m+1}\hat H_{m+1}(v)e^{-v^2/2} + \int_{-\infty}^\infty \mathrm{d}v\,  f_k(t,v)\sqrt{m}\hat H_{m-1}(v)e^{-v^2/2}\right] \notag \\
    = ik \left( \sqrt{m+1} \tilde f(t,m+1) + \sqrt{m} \tilde f(t,m-1) \right).
\end{align}
Hence, Eq.~\ref{eq:hermite-vfp} reduces to
\begin{equation}
    \partial_t \tilde f(t,m) + ik\left(\sqrt{m+1}\tilde f(t,m+1) + \sqrt{m}\tilde f(t,m-1)\right) = 0.
    \label{eq:recurrence-relation}
\end{equation}
For $m\gg1$, we treat $m$ as a real number that can take continuous values and $\tilde f(t,m)$ as a smooth function of both $t$ and $m$, an approximation rigorously justified in previous work \citep[see][]{kanekar_fluctuation-dissipation_2015, schekochihin_phase_2016}. Expanding $\tilde f(t,m\pm1)$ about $m$ using a Taylor series ($\tilde f(t,m\pm1) \approx \tilde f(t,m) \pm \partial_m \tilde f(t,m)$) and substituting into Eq.~\ref{eq:recurrence-relation} gives
\begin{align}
    \partial_t \tilde f(t,m) + ik\left[(\sqrt{m+1} + \sqrt{m})\tilde f(t,m) + (\sqrt{m+1} - \sqrt{m})\partial_m\tilde f(t,m)\right] \approx 0.
\end{align}
Since $\sqrt{m+1} \approx \sqrt{m}$ to leading order, the previous equation simplifies to
\begin{equation}
    \partial_t \tilde f(t,m) + 2ik\sqrt{m}\tilde f(t,m) + \frac{ik}{2\sqrt{m}}\partial_m \tilde f(t,m) \approx 0.
    \label{eq:hermite-pde}
\end{equation}
The first two terms describe the rapid oscillation of $\tilde f$ with frequency $2k\sqrt{m}$. To isolate the cascade dynamics, we remove the oscillatory component by assuming $\tilde f(t,m)$ is the product of a slowly varying envelope and a complex phase term,
\begin{equation}
    \tilde f(t,m) = g(t,m) e^{-2ik\sqrt{m}t}.
\end{equation}
Substituting this into Eq.~\ref{eq:hermite-pde} yields
\begin{equation}
    \partial_t g(t,m) + \frac{ik}{2\sqrt{m}} \partial_m g(t,m) + \frac{k^2t}{2m}g(t,m) \approx 0.
    \label{eq:pre-advect-eq}
\end{equation}
Since $m\gg1$, the $1/m$ term is marginal and can be neglected to leading order, yielding
\begin{equation}
    \partial_t g(t,m) + \frac{ik}{2\sqrt{m}} \partial_m g(t,m) \approx 0.
\end{equation}
This is an advection equation in $m$ with characteristic speed
\begin{equation}
    \frac{dm}{dt} \propto \frac{k}{2\sqrt{m}}.
    \label{eq:characteristic-speed}
\end{equation}
We consider the local cascade time at a given $m$: how long it takes for information to cascade from mode $m$ to $m+1$,
\begin{equation}
    t_{\rm cas}(m) \propto \frac{(m+1) - m}{dm/dt} = \frac{2\sqrt{m}}{k},
    \label{eq:cascade-time-hermite}
\end{equation}
As in our derivation in Section~\ref{sec:linear-fourier-mode}, the next step is to determine $t_{\rm coll}(m)$ from the Lenard-Bernstein operator (Eq.~\ref{eq:LB}). Since the Hermite modes are eigenfunctions of the Lenard--Bernstein operator, the VFP equation (Eq.~\ref{eq:VFP}) gives (neglecting streaming)
\begin{align}
    \partial_t \tilde f(t,m) \propto C[\tilde f(t,m)] = -m\nu \tilde f(t,m), \\
    \tilde f(t,m) \propto \tilde f(0,m)e^{-m\nu t}.
\end{align}
Hence, the collisional damping rate and time are given by
\begin{align}
    \gamma_{\rm coll} \propto m\nu, \\
    t_{\rm coll}(m) \propto \frac{1}{m\nu}. \label{eq:collision-time-hermite}
\end{align}
Equating Eq.~\ref{eq:cascade-time-hermite} and Eq.~\ref{eq:collision-time-hermite} gives the critical Hermite mode
\begin{equation}
    m_c \propto \left( \frac{k}{\nu}\right)^{2/3}.
    \label{eq:critical-hermite-linear-before-sub}
\end{equation}
Making the same substitutions as in Eq.~\ref{eq:critical-wave-number-2}, we rewrite Eq.~\ref{eq:critical-hermite-linear-before-sub} as
\begin{equation}
    m_c \propto \left( \frac{k\lambda_D}{ \hat \nu}\right)^{2/3}.
    \label{eq:crit-hermite-linear}
\end{equation}

\subsection{Nonlinear critical Fourier and Hermite modes}
\label{sec:nonlinear-mode-scaling}

In nonlinear Landau damping, the electric field amplitude is large enough such that charged particles moving near the phase velocity of the EPW become trapped in its potential well. Particle trapping manifests as a vortex in phase space (see figure~\ref{fig:phase-space}), and particles within the vortex oscillate according to their bounce frequency; \citet{oneil_collisionless_1965} showed that the corresponding trapping timescale is
\begin{equation}
    t_{\rm tr} \propto \frac{1}{\omega_b},
    \label{eq:tr-time}
\end{equation}
where $\omega_b$ is the dimensionless bounce frequency normalised to $\omega_p$. Particles outside of the vortex are not trapped (passing) and undergo velocity modulation as they traverse the periodic potential. This causes the separatrix layer between trapped and passing particles to develop fine gradients with respect to velocity, resulting in nonlinear phase mixing. Nonlinear phase mixing associated with particle trapping replaces filamentation as the dominant mechanism generating fine velocity-space structure at the point when when the two processes occur on the same time scale ($t_{\rm cas} \sim t_{\rm tr}$). In the absence of collisions, the separatrix width becomes infinitesimally thin \citep{joglekar_evolution_2026}, leading to infinitely large gradients in velocity; as was the case in linear Landau damping, this process corresponds to the velocity-space cascade as the system evolves towards so-called Bernstein-Greene-Kruskal modes \citep{bernstein_exact_1957}.

Collisional diffusion necessarily smooths fine gradients in phase space, halting nonlinear phase mixing and the cascade to higher velocity-space modes. The nonlinear critical mode is the point at which this arrest occurs, equivalent to the mode required to resolve the separatrix width. To determine the nonlinear critical mode, we equate the trapping time (Eq.~\ref{eq:tr-time}) with the collisional timescale expressed in the Fourier and Hermite basis (Eq.~\ref{eq:damping-time} and Eq.~\ref{eq:collision-time-hermite}). For the nonlinear critical Fourier mode, we obtain
\begin{equation}
    \overline s_{c} \propto \sqrt{\frac{\omega_b}{\hat \nu}},
    \label{eq:nonlinear-fourier}
\end{equation}
where the overline denotes a nonlinear critical value. Similarly, for the Hermite representation, we obtain
\begin{equation}
    \overline m_c \propto \frac{\omega_b}{\hat \nu}.
    \label{eq:nonlinear-hermite}
\end{equation}

\section{Model and Diagnostics}
\label{sec:model}

Landau damping is simulated using \texttt{ADEPT} \citep{joglekar_machine_2023}, which uses a semi-Lagrangian scheme to solve the VFP equation (Eq.~\ref{eq:VFP}) and the dimensionless Poisson's equation, $\partial_x E = 1 - \int \mathrm{d}v\, f$, for the distribution function $f(t,x,v)$ and electric field $E(t, x)$ with periodic boundary conditions. In these simulations, $N_x = 128$, $N_v = 4096$, $L_x = 2\pi/k\lambda_D$, $\Delta t = 0.2$ $\omega_p^{-1}$, and $t_{\max} = 2000$ $\omega_p^{-1}$. In each simulation an EPW is excited by a finite-time external driver of strength $a_0$; after the driver is switched off the wave evolves freely and undergoes Landau damping. The driving protocol is described fully in \citet{joglekar_machine_2023}.

We perform a parameter scan over $a_0$, $k\lambda_D$, and $\nu$. The scan consists of eight values of $k\lambda_D \in \{0.26, 0.28, \dots, 0.40\}$, and ten logarithmically spaced values each of $a_0 \in \{10^{-7}, 10^{-3}\}$ and $\nu \in \{10^{-6}, 10^{-3}\}$. For each combination of parameters, an independent simulation is performed, resulting in a total of $8 \times 10 \times 10 = 800$ simulations. The ranges in $a_0$, $k\lambda_D$, and $\nu$ are chosen to capture linear and nonlinear Landau damping.

We seek to numerically extract the critical mode from the distribution function obtained in each \texttt{ADEPT} simulation result. To isolate the wave dynamics, we decompose the distribution function $f(t, x, v)$ using a spatial Fourier series. The first spatial mode $f_{k_1}(t,v)$ corresponding to the fundamental wavenumber $k_1 = 2\pi/L_x$ is calculated via
\begin{equation}
    f_{k_1}(t,v) = \frac{1}{L_x}\int_0^{L_x}\mathrm{d}x\,f(t,x,v)e^{-ik_1 x}.
    \label{eq:fft-in-space}
\end{equation}
We then perform Fourier and Hermite transforms in velocity on $ f_{k_1}(t,v)$. The Hermite transform (Eq.~\ref{eq:hermite-transform-operator}) yields the Hermite coefficients as a function of $t$ and $m$,
\begin{equation}
    \tilde f_{k_1}(t,m) = \frac{1}{\sqrt{2\pi}}\int_{-\infty}^{\infty} \mathrm{d}v\,f_{k_1}(t,v)\hat H_m (v)e^{-v^2/2}.
    \label{eq:hermite-transform}
\end{equation}
To numerically evaluate Eq.~\ref{eq:hermite-transform}, we rewrite it as
\begin{equation}
    \tilde f_{k_1}(t,m) = \frac{1}{\sqrt{2\pi}}\int_{-\infty}^{\infty}\mathrm{d}v\,e^{-v^2}\left[  f_{k_1}(t,v)\hat H_m (v)e^{v^2/2} \right],
\end{equation}
which is suitable for Gauss-Hermite quadrature approximation, giving
\begin{equation}
    \tilde f_{k_1}(t, m) \approx \frac{1}{\sqrt{2\pi}}\sum_{i=1}^M w_i \hat H_m(v_i)  f_{k_1}(t,v_i)e^{v_i^2/2},
    \label{eq:gauss-hermite-quadrature}
\end{equation}
where $w_i$ are the standard probabilist's Gauss-Hermite quadrature weights evaluated at the roots $v_i$ of the $M$th-order Hermite polynomial. We compute Eq.~\ref{eq:gauss-hermite-quadrature} using the Julia module \texttt{FastGaussQuadrature.jl} \citep{townsend_fast_2016}.

\begin{figure*}
    \centering
    \includegraphics[width=\textwidth]{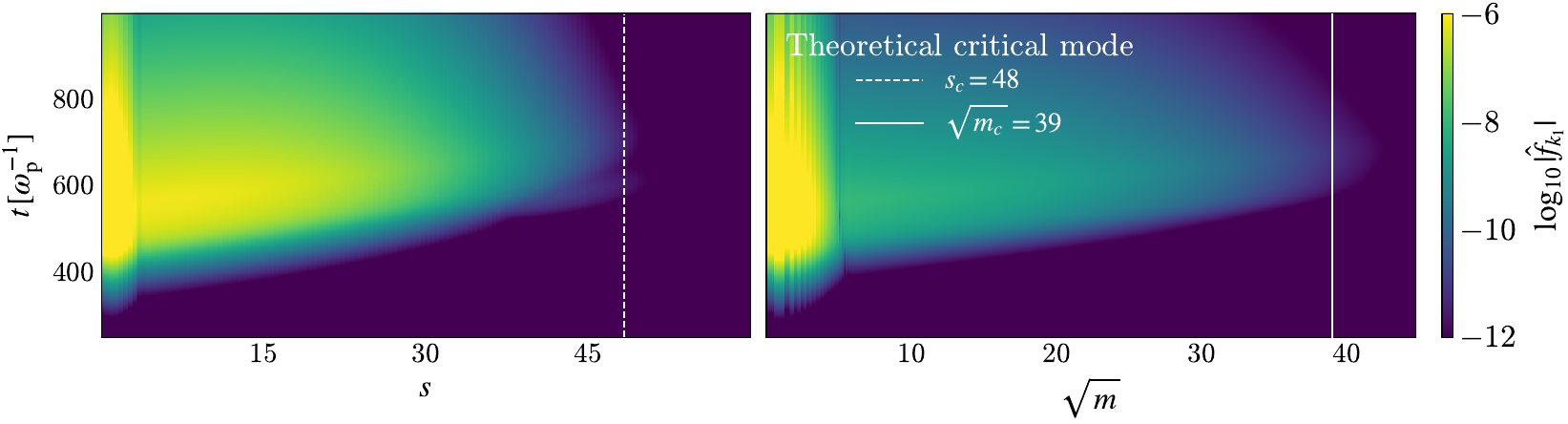}
    \caption{Velocity-space cascade for a representative linear case ($\omega_b = 0.009$, $k\lambda_D = 0.30$, $\hat{\nu} = \SI{1.0e-4}{}$). \textit{Left}: $\log_{10}|\hat{f}_{k_1}(t, s)|$ as a function of time and the dimensionless Fourier velocity wavenumber $s$. \textit{Right}: $\log_{10}|\hat{f}_{k_1}(t, m)|$ as a function of time and the square root of the Hermite mode number. In both panels, the cascade front advances linearly to higher mode numbers as phase mixing proceeds, before being arrested by collisional diffusion. The white lines mark the theoretical critical mode ($s_c$, dashed, left; $\sqrt{m_c}$, solid, right) from Eqs.~\ref{eq:critical-wave-number-2} and~\ref{eq:crit-hermite-linear}. The coefficients are determined from the linear least-squares fit (see table~\ref{tab:coefficient_fits}).}
    \label{fig:cascade-linear}
\end{figure*}

\begin{figure*}
    \centering
    \includegraphics[width=\textwidth]{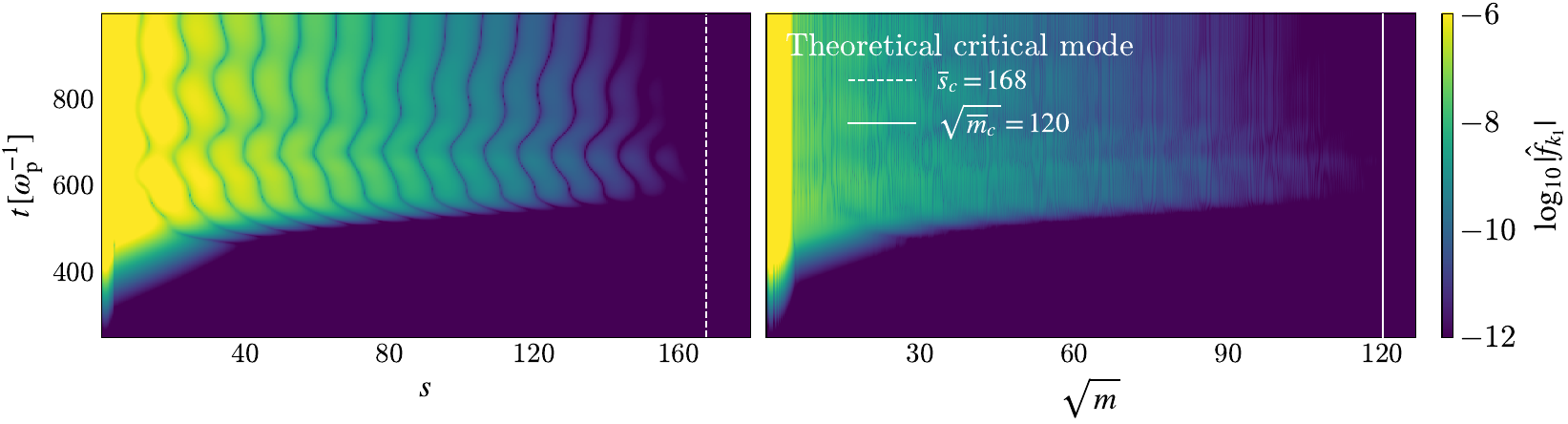}
    \caption{Same as Figure~\ref{fig:cascade-linear}, but for a representative nonlinear case ($\omega_b = 0.049$, $k\lambda_D = 0.28$, $\hat{\nu} = \SI{2.15e-5}{}$) in which particle trapping, instead of filamentation, drives the cascade to higher modes. Bounce oscillations are visible as periodic recurrences in the cascade front, reflecting the closed-loop recirculation of trapped particles. The white lines indicate the theoretical critical mode ($\overline s_c$, dashed, left; $\sqrt{\overline m_c}$, solid, right), computed from Eqs.~\ref{eq:nonlinear-fourier} and~\ref{eq:nonlinear-hermite} using the fitted coefficients in table~\ref{tab:coefficient_fits}.}
    \label{fig:cascade-nonlinear}
\end{figure*}

We implement a numerical algorithm to determine the critical mode from the transformed primary spatial perturbation $\hat f_{k_1}$ (the hat accent denoting either a Hermite or Fourier transform). The velocity mode number at which the information cascade terminates is precisely the critical mode; $\hat f_{k_1}$ captures this behaviour in velocity-space, as can be seen in Figures~\ref{fig:cascade-linear} and~\ref{fig:cascade-nonlinear} for the linear and nonlinear regimes, respectively. To determine the critical mode, the algorithm analyses slices of $\tilde f_{k_1}(t, n)$ in $n$, where $n$ is the mode index denoting either $s$ or $m$. The critical mode is defined as
\begin{equation}
    n_{c} \equiv \min \left\{\, n \;\middle|\; \lvert \hat f_{k_1}(t,n) \rvert < \epsilon \ \text{for all } t \right\},
    \label{eq:n-crit}
\end{equation}
where $\epsilon= 10^{-12}$ is the noise threshold we use in this work.

After extracting the critical Fourier and Hermite modes across the parameter space, we partition the simulations into linear and nonlinear regimes according to the nonlinearity parameter $\omega_b/\gamma_L$ and collisionality parameter $\omega_b / \hat \nu$, where $\gamma_L$ is the linear Landau damping rate. Simulations with $\omega_b/\gamma_L<1$ are classified as linear (638/800 simulations) while those with $\omega_b/\gamma_L>5$ and $\omega_b / \hat \nu > 25$ are classified as nonlinear (39/800). We classify the simulations in this manner so that we only consider the purely linear and nonlinear regimes. The remaining 123 simulations fall into an intermediate, semi-nonlinear regime and are excluded; the analysis of this regime lies outside the scope of this paper.

\section{Results}
\label{sec:results}

\begin{table*}
    \centering
    \begin{tabularx}{0.25\textwidth}{l C}
    \hline
        Quantity & $C$ \\
        \hline
        $s_c$ (linear)             & $3.4$ \\
        $m_c$ (linear)               & $7.3$ \\
        $\overline{s}_{c}$ (nonlin.) & $3.5$ \\
        $\overline{m}_c$ (nonlin.)   & $6.4$ \\
    \hline
    \end{tabularx}
    \caption{Results from ordinary least squares fit to Eq.~\ref{eq:fit-eq}, with $C$ as the only free parameter. Four separate critical mode fits were performed: linear Fourier $s_c$ ($R^2 = 0.95$), linear Hermite $m_c$ ($R^2 = 0.86$), nonlinear Fourier $\overline s_c$ ($R^2 = 0.97$), and nonlinear Hermite $\overline m_c$ ($R^2=0.90$). Simulations in which the cascade exceeded the number of available modes were excluded from the fitting procedure.}
    \label{tab:coefficient_fits}
\end{table*}

\begin{table*}
  \centering
  \begin{tabularx}{\textwidth}{l C CC CC CC}
    \hline
    & & \multicolumn{2}{c}{$\alpha_{\omega_b}$} 
      & \multicolumn{2}{c}{$\beta_{k\lambda_D}$} 
      & \multicolumn{2}{c}{$\gamma_{\hat\nu}$} \\
    \cmidrule(lr){3-4}\cmidrule(lr){5-6}\cmidrule(lr){7-8}
    Quantity & $C$ & theory & fit & theory & fit & theory & fit \\
    \hline
    $s_c$ (linear)             & $7.7$      & $0$   & $0.10$ & $1/3$ & $0.61$ & $-1/3$ & $-0.34$ \\
    $m_c$ (linear)               & $366.3$    & $0$   & $0.31$ & $2/3$ & $2.48$ & $-2/3$ & $-0.66$ \\
    $\overline{s}_{c}$ (nonlin.) & $10.2$     & $1/2$ & $0.40$ & $0$   & $0.51$ & $-1/2$ & $-0.43$ \\
    $\overline{m}_c$ (nonlin.)   & $1684.0$   & $1$   & $0.82$ & $0$   & $2.86$ & $-1$   & $-0.77$ \\
    \hline
  \end{tabularx}
  \caption{Same as table~\ref{tab:coefficient_fits}, but with $C$, $\alpha$, $\beta$, and $\gamma$ treated as free parameters. The theoretical values derived in section~\ref{sec:critical-mode-scaling} are shown alongside the fitted values. The $R^2$ values for $s_c$, $m_c$, $\overline s_c$, and $\overline m_c$ are 0.99, 0.97, 0.99, and 0.99, respectively.}
  \label{tab:power_law_fits}
\end{table*}

\begin{figure*}
    \centering
    \includegraphics[width=\textwidth]{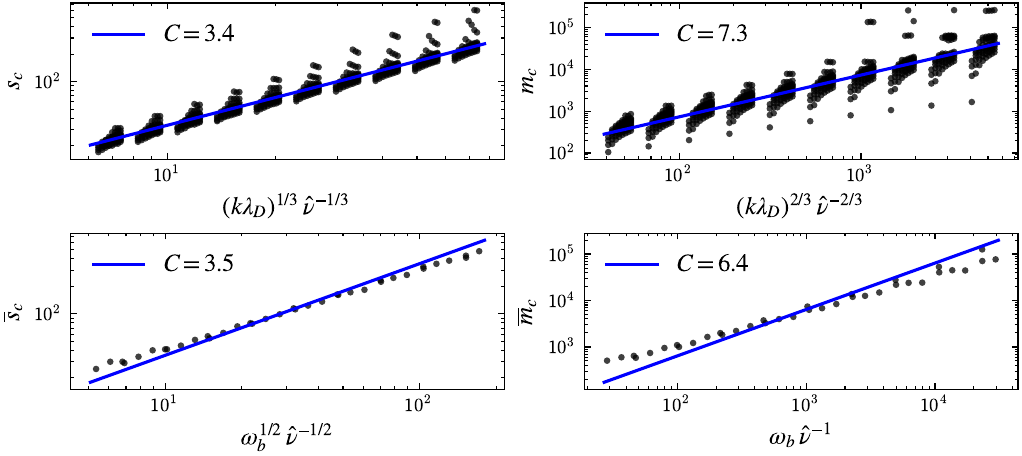}
    \caption{Measured critical mode $n_c$ versus the theoretical prediction on log-log axes. The best-fit line (blue) corresponds to Eq.~\ref{eq:fit-eq} evaluated using the best-fit $C$ with fixed theoretical exponents. The four panels show, clockwise from top-left: $s_c$ (linear Fourier), $m_c$ (linear Hermite), $\overline{m}_c$ (nonlinear Hermite), and $\overline{s}_{c}$ (nonlinear Fourier). The corresponding $R^2$ values are given in table~\ref{tab:coefficient_fits}. Simulations where the cascade exceeded the number of available modes were excluded from each plot.}
    \label{fig:scaling-verification}
\end{figure*}

We perform two sets of ordinary least squares fits on the linear and nonlinear Fourier and Hermite critical modes of the form
\begin{equation}
    n_c = C \omega_b^\alpha (k\lambda_D)^\beta \hat \nu^\gamma,
    \label{eq:fit-eq}
\end{equation}
where $n_c$ is the critical mode, $C$ is a dimensionless proportionality constant, and $\alpha$, $\beta$, and $\gamma$ are scaling exponents associated with bounce frequency, wavenumber, and collisional frequency, respectively. The first set fits only $C$, fixing $\alpha$, $\beta$, and $\gamma$ to their theoretical values determined in section~\ref{sec:critical-mode-scaling}, and is summarised in table~\ref{tab:coefficient_fits}. The second set fits $C$ along with the exponents $\alpha$, $\beta$, and $\gamma$, and is summarised in table~\ref{tab:power_law_fits}.

\subsection{Fixed-exponent fits}
\label{sec:Fixed-exponent fits}

The fixed-exponent Fourier fits in table~\ref{tab:coefficient_fits} yield $R^2$ values greater than $0.95$, indicating excellent agreement with theory. The Hermite fits yield $R^2>0.86$, showing less, but still reasonable, agreement with theory. This difference is primarily associated with $\beta$; the Fourier fits obey the theoretical $\beta$ scaling better than their Hermite counterparts. This becomes evident when we fit the exponents (see table~\ref{tab:power_law_fits}). Figure~\ref{fig:scaling-verification} shows that the fitted coefficients provide accurate prefactors to the theoretical scaling laws. In addition, we expect the Hermite coefficients to be twice the value of their Fourier counterparts \citep{boyd_asymptotic_1984}, which is roughly what we observe. These results suggest that the fitted prefactors can be useful for spectral solvers in estimating the mode number required to accurately resolve the velocity-space cascade.

\subsection{Free-exponent fits}
\label{sec:free-exponent-fits}

The linear fits $s_c$ and $m_c$ in table~\ref{tab:power_law_fits} recover non-zero values of $\alpha$ despite fitting within the linear regime where theory predicts $\alpha=0$. Since $\omega_b \propto \sqrt{\phi_0}$, the $\phi_0$ dependency in these fits is actually no larger than $\phi_{0}^{\alpha/2}$. This implies that the linear mode numbers show only a very weak dependence on $\phi_0$: $s_c \propto \phi_0^{0.05}$ and $m_c\propto \phi_0^{0.16}$. Overall, the $s_c$ and $m_c$ fits accurately recover the theoretical behaviour in the $\omega_b$ scaling. In addition, these fits recover $\gamma$ within $1.94\%$ and $1.26\%$ of their respective theoretical values, indicating strong agreement with theory.

The $s_c$ and $m_c$ fits fail to accurately recover $\beta$, as do their nonlinear counterparts, $\overline s_c$ and $\overline m_c$. Given that the Landau damping rate depends strongly on $k\lambda_D$, the scan in $k\lambda_D$ values has a relatively small sample size and sample rate. A more thorough parameter scan would likely improve the recovered $\beta$ values.

The nonlinear Fourier fit $\overline s_c$ recovers $\alpha$ and $\gamma$ within $19.33\%$ and $14.05\%$ of their theoretical values, respectively, indicating reasonable agreement with theory. The nonlinear Hermite fit $\overline m_c$ shows deviations of $18.19\%$ and $23.32\%$ in $\alpha$ and $\gamma$. The discrepancies in the Hermite values reflect the numerical challenge of performing an accurate Hermite transform at high modes, specifically the sensitivity of the Hermite expansion to the tails of the distribution function. The Gaussian weight $e^{v_i^2/2}$ in the Gauss-Hermite quadrature (Eq.~\ref{eq:gauss-hermite-quadrature}) exponentially amplifies regions of large $|v_i|$, where the interpolation of $f$ onto the quadrature abscissas introduces the largest absolute error. Near the separatrix, sharp gradients in velocity-space further reduce interpolation accuracy. The nonlinear Fourier fit, by contrast, does not involve this exponential amplification and operates directly in a bounded frequency domain, making it more robust to localized velocity-space errors. Moreover, since the Hermite cascade scales as the square of the Fourier cascade \citep{boyd_asymptotic_1984}, the reasonable agreement of the nonlinear Fourier fit $\overline s_c$ provides indirect validation that discrepancies in $\overline m_c$ are primarily numerical.

The Hermite fits yield significantly larger prefactors than those obtained from the fixed-exponent fits in table~\ref{tab:coefficient_fits}. This behaviour is likely a consequence of inaccuracies observed in $\beta$: allowing $\beta$ to take values substantially larger than its theoretical prediction shifts part of the scaling dependence into the prefactor, resulting in artificially inflated values of $C$. Although the Fourier fits also exhibit deviations in $\beta$, these are considerably smaller, and the corresponding prefactors remain closer to those obtained in the fixed-exponent fits.

As mentioned in section~\ref{sec:Fixed-exponent fits}, the critical Hermite mode coefficient is expected to be twice that of the critical Fourier mode coefficient. The fixed-exponent fits are consistent with this expectation, whereas the free-exponent Hermite fits are not. For this reason, the prefactors reported in table~\ref{tab:coefficient_fits} provide more reliable estimates for practical numerical applications.

\section{Conclusion}
\label{sec:conclusion}

\begin{table*}
    \centering
    \begin{tabularx}{1.0\textwidth}{l C C C}
    \hline
        & & \multicolumn{2}{c}{Hermite modes required} \\
        \cmidrule(lr){3-4}
        Regime & $\hat{\nu}$ & Linear damping & Nonlinear damping\\
        \hline
        Interstellar medium  & \SI{1.17e-9}{} & \SI{3e6}{} -- \SI{4e6}{} & \SI{2e7}{} -- \SI{3e8}{} \\
        Solar corona         & \SI{3.0e-8}{}  & \SI{3e5}{} -- \SI{4e5}{} & \SI{8e5}{} -- \SI{1e7}{} \\
        Hot plasma           & \SI{6.67e-6}{} & 8401 -- \SI{1e4}{} & 3534 -- \SI{5e4}{} \\
        Theta pinch          & \SI{5.0e-5}{}  & 2192 -- 2922 & 471 -- 6417 \\
        Dense hot plasma     & \SI{3.33e-4}{} & 619 -- 825 & 71 -- 963 \\
        Laser plasma         & \SI{3.33e-3}{} & 133 -- 178 & 7 -- 96 \\
    \hline
    \end{tabularx}
    \caption{Approximate number of Hermite modes required to accurately resolve
    linear and nonlinear Landau damping in different plasma regimes. These estimates represent upper bounds and do not necessarily correspond to the minimum resolution required to accurately reproduce macroscopic behavior. We calculated
    the mode numbers using Eqs.~\ref{eq:crit-hermite-linear}
    and~\ref{eq:nonlinear-hermite} along with the respective prefactors in Table~\ref{tab:coefficient_fits}. Collisional frequency values were obtained from the NRL Plasma Formulary. We present ranges of mode numbers based on the range of wave numbers observed in the linear simulations ($k\in\{0.26,0.40\}$) and the bounce frequencies observed in the nonlinear simulations ($\omega_b\in\{0.004,0.050\}$).}
    \label{tab:plasma_regimes}
\end{table*}

In this work, we have derived analytical scalings for the critical Fourier and Hermite velocity-space mode numbers in both linear and nonlinear Landau damping for the 1D-1V Vlasov--Poisson system. These scalings, given by Eqs.~\ref{eq:critical-wave-number-2},~\ref{eq:crit-hermite-linear},~\ref{eq:nonlinear-fourier}, and~\ref{eq:nonlinear-hermite}, were obtained using a unified cascade-balance argument, in which the characteristic time scale of collisional diffusion is equated with phase mixing in the linear regime and particle trapping in the nonlinear regime. Following this framework in the Fourier and Hermite basis allowed us to extract power-laws for the critical mode in terms of $\omega_b$, $k\lambda_D$, and $\hat\nu$.

We compared these scalings against an ensemble of 800 \texttt{ADEPT} simulations spanning linear and nonlinear regimes of Landau damping. We only considered simulations exhibiting purely linear or nonlinear damping, disregarding intermediate, semi-nonlinear simulations. We performed an ordinary least squares fit on the data and accurately recovered the $\omega_b$ and $\hat\nu$ scalings. By fixing the theoretical exponents and fitting for the prefactors, we obtain quantitative estimates for the critical mode thresholds in both Fourier and Hermite space. These results provide an upper bound on the number of modes needed to completely resolve Landau damping given known plasma conditions. 

We highlight the approximate range of Hermite modes needed to resolve nonlinear Landau damping for plasmas belonging to select regimes in Table~\ref{tab:plasma_regimes}. In the weakly collisional regimes, the number of Hermite modes required is impractically large for use in spectral solvers. In future work, we hope to characterise the minimum spectral resolution requirement, i.e. the velocity-space mode number beyond which information can be safely discarded while preserving the important macroscopic behaviour of the system. This will provide a more practical heuristic to inform continuum Vlasov codes.

This material is based upon work supported by IFE COLoR and IFE-STAR under U.S. Department of Energy Grant No. DE-SC0024863 and DE-SC0025573. This research used resources of the National Energy Research Scientific Computing Center, a DOE Office of Science User Facility supported by the Office of Science of the U.S. Department of Energy under Contract No. DE-AC02-05CH11231 using NERSC award FES-ERCAP0026741. A. J. thanks useful discussions with B. Afeyan to motivate this study. A. J. and J. C. also acknowledge funding from Pasteur Labs.

\bibliographystyle{jpp}

\bibliography{ortho-basis}

\end{document}